\documentclass[preprintnumbers,aps,prd,twocolumn,showpacs,showkeys,floatfix,preprintnumbers,letterpaper,amsmath,amssymb,superscriptaddress]{revtex4-1}
\usepackage{subfigure}
\usepackage{graphicx}
\usepackage{dcolumn}
\usepackage{epsfig}
\usepackage{amsmath}
\usepackage{amsfonts}
\usepackage{amssymb}
\usepackage{color} 
\usepackage{xcolor}
\usepackage{hyperref}
\usepackage{tabularx}
\usepackage{float}

\newcommand{\mnras}{Mon. Not. R. Astron. Soc.}
\newcommand{\aap}{A\&A}

\newcommand{\apjl}{Astrophys. J. Lett.}

\usepackage[T1]{fontenc}
\usepackage{ae,aecompl}
\newcommand{\be}{\begin{equation}}
\newcommand{\ee}{\end{equation}}
\newcommand{\bea}{\begin{eqnarray}}
\newcommand{\eea}{\end{eqnarray}}

\newcommand{\bx}{{\bf x}}

\newcommand{\bv}{{\bf v}}

\newcommand{\bq}{{\bf q}}
\newcommand{\bS}{{\bf S}}

\def\bx{{\bf x}}         
\def\bq{{\bf q}}         
\def\bv{{\bf v}}         

\begin{document}

\title{Optimal Transport Reconstruction of Biased Tracers in Redshift Space}

\author{Farnik Nikakhtar}
\email{farnik.nikakhtar@yale.edu}
\affiliation{Department of Physics, Yale University, New Haven, CT 06511, USA}

\author{Nikhil Padmanabhan}
\affiliation{Department of Physics, Yale University, New Haven, CT 06511, USA} 
\affiliation{Department of Astronomy, Yale University, New Haven, CT 06511, USA}

\author{Bruno L\'evy}
\affiliation{Centre Inria de Paris, 2 Rue Simone Iff, 75012 Paris, France}

\author{Ravi K.~Sheth}
\affiliation{Center for Particle Cosmology, University of Pennsylvania, Philadelphia, PA 19104, USA}
\affiliation{The Abdus Salam International Center for Theoretical Physics, Strada Costiera 11, Trieste 34151, Italy}
    
\author{Roya Mohayaee}
\affiliation{Sorbonne Universit\'e, CNRS, Institut d'Astrophysique de Paris, 98bis Bld Arago, 75014 Paris, France}
\affiliation{Rudolf Peierls Centre for Theoretical Physics, University of Oxford, Parks Road, Oxford OX1 3PU, United Kingdom}

\date{\today}

\begin{abstract}

Recent research has emphasized the benefits of accurately reconstructing the initial Lagrangian positions of biased tracers from their positions at a later time, to gain cosmological information. A weighted semi-discrete optimal transport algorithm can achieve the required accuracy, provided the late-time positions are known, with minimal information about the background cosmology. The algorithm's performance relies on knowing the masses of the biased tracers, and depends on how one models the distribution of the remaining mass that is not associated with these tracers. We demonstrate that simple models of the remaining mass result in accurate retrieval of the initial Lagrangian positions, which we quantify using pair statistics and the void probability function. This is true even if the input positions are affected by redshift-space distortions. The most sophisticated models assume that the masses of the tracers, and the amount and clustering of the missing mass are known; we show that the method is robust to realistic errors in the masses of the tracers and remains so as the model for the missing mass becomes increasingly crude.

\end{abstract}


\maketitle


\section{Introduction}\label{intro}

The baryon acoustic oscillation (BAO) feature in the two-point statistics of
galaxies and gas provides a ruler to constrain the expansion history
of the Universe.  However, this ruler must be standardized, to undo the effects
of bulk flows between the time of last scattering and when the galaxies were
observed. Most density field reconstruction methods which seek to do this make
assumptions about the background cosmology \cite{recSDSS, recIterate, recHE,
eFAM2019}.  An alternative approach based on Optimal Transport (OT)\cite{nature,EUR, royaMAK} simply assumes (a) that the initial density fluctuation field was uniform (i.e. structure growth is seeded by inhomogeneities in the {\em
displacement} field), and (b) that the displacements which map initial to final
positions (or vice versa) can be described by a convex potential.  
These are fairly generic assumptions which do not require one to know any other
details about the background cosmology.

A discrete OT algorithm, with limited computational power, has already been applied to haloes and real data in redshift space to retrieve the cosmic density and velocity fields and constrain cosmological parameters \citep{AA,brentapj1,mathis,brentapj2}. However, to reconstruct the BAO features, far more powerful OT algorithms adequate for present big data is needed. Recently,
we have developed a powerful new semi-discrete OT algorithm and
have shown that for dark matter in real space, it performs
extremely well, meaning that it reconstructs the BAO feature to subpercent precision \cite{PRLdm}.
Here we further develop our algorithm for application to biased tracers in redshift space, and provide a complete numerical framework for BAO reconstruction of existing and forthcoming redshift surveys. 

For biased tracers, such as halos, the method requires a
reasonable estimate both of the masses of the tracers as well as of the distribution
of the mass that is not directly associated with the tracers \cite[which,
following Ref.][we refer to as `dust']{PRLhalos}.  This is
because assumption (a) is crucial to the speed of the algorithm, since it is only
the {\em total} matter field which was uniform initially:  the initial locations
of biased tracers can be very non-uniform (e.g. it depends on halo mass)
\citep{mw1996, st1999}.
The first goal of this paper is to develop and test a simple model for the dust which 
 (a) can be generated based solely on the observed locations of the biased tracers, and 
 (b) when incorporated into the OT analysis, enables a sufficiently accurate reconstruction of the initial field.  
 This is the subject of section~\ref{sec:dust}; details about our fiducial dust
 model as well as alternatives we considered are provided in
 Appendix~\ref{app:U}.

\begin{figure*}[ht]
    \centering
    \includegraphics[width=\textwidth]{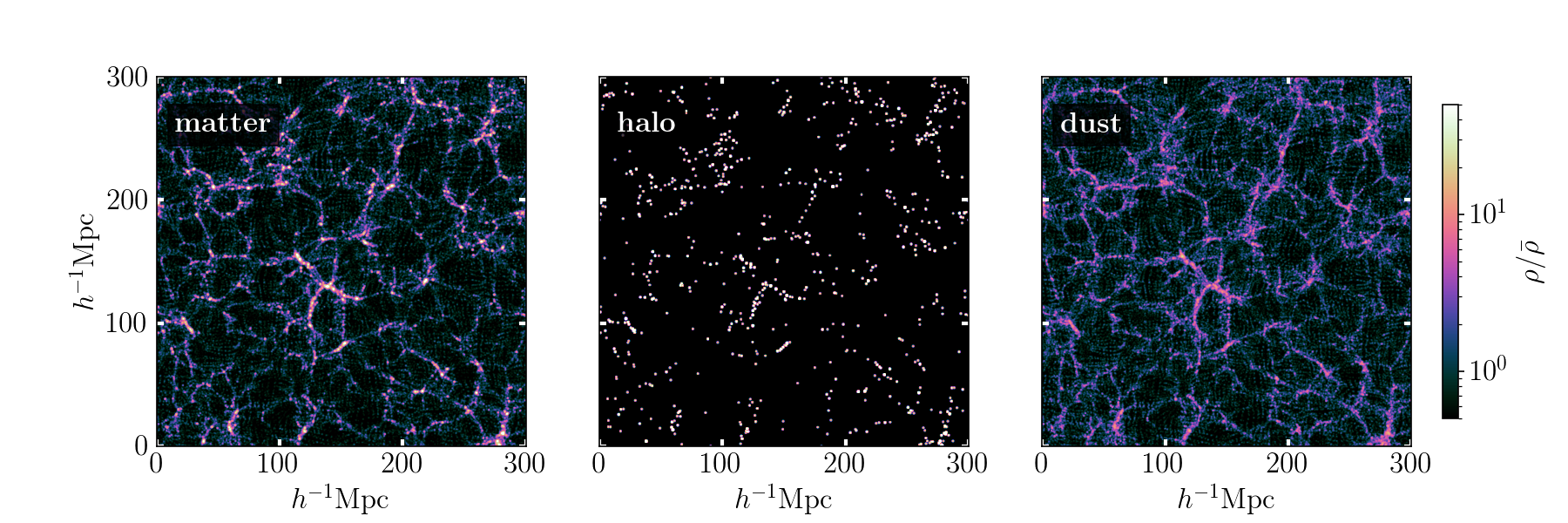}
    \caption{Projected dark matter (left), halo (middle), and dust (right) density fields at $z=0$ of the HADES simulations \cite{hades} in the same $300\times300\times 20 h^{-3}{\rm Mpc}^3$ comoving volume. 
    \label{fig:fields}}
\end{figure*}

Our second goal (Sec.~\ref{sec:zspace}) is to investigate whether our dust and OT framework can be extended to treat the (more realistic) case in which the Eulerian positions are not given in real space, but are instead redshift-space distorted.  In section~\ref{sec:zspace}, we motivate and introduce an {\em anisotropic}
semi-discrete optimal transport algorithm and demonstrate its effectiveness in
accurately reconstructing initial Lagrangian positions even in the presence of
redshift-space distortions.

We illustrate our results using measurements in the HADES simulations
\cite{hades} of a flat $\Lambda$CDM background cosmology that has
$(\Omega_m,\Omega_b,h)=(0.3175,0.049,0.6711)$, since they were also used in our
real-space OT study \cite{PRLhalos}. The simulations follow the gravitational evolution of
$512^3$ identical particles in a periodic box of side $L=1h^{-1}$Gpc, from an
initially Gaussian fluctuation field having power spectrum $P_{\rm Lin}(k)$ with
shape and amplitude parameters $(n_s,\sigma_8)=(0.9624,0.833)$.  Halos were
identified in the $z=0$ output of each box using a friends-of-friends algorithm
with linking length that is $0.2\times$ the inter-particle spacing.  Our
fiducial sample includes all halos with more than 20 particles, i.e. more
massive than $20m_p$ where $m_p=6.5\times 10^{11}h^{-1}M_\odot$.  This sample
accounts for $\sim 22$ percent of the mass in a simulation box.  However, to
explore the dependence on mass fraction and clustering strength of the biased
sample, we also show some results from a high and low mass sample, with mass
cuts chosen to result in a mass fraction of $\sim 10$ percent (see
Table~\ref{tab:1}).  
In {\em all} the results which follow, we weight each halo
by its mass when computing e.g. the density field or correlation functions. 
Sec.~\ref{sec:LNmass} discusses the impact of mass errors on our results.

\begin{table}
  \centering
  \begin{tabular}{|c|c|c|c|c|c|}
  \hline
    Case & Mass [$10^{13}M_\odot/h$] & $p$ & $b_{h}$ & $b_{d}$\\
    \hline \hline 
    $r$-space all & $1.3 < m_h$ & 0.22 & 2.08 & 0.68\\
    \hline
    $r$-space mid & $1.3 < m_h < 7$ & 0.11 & 1.42 & 0.94\\
    \hline
    $r$-space high & $10 < m_h$ & 0.10 & 3.19 & 0.78\\
    \hline\hline
    $z$-space all & $1.3 < m_h$ & 0.22 & 2.25 & 0.86\\
    \hline
  \end{tabular}
  \caption{The halo samples used in this paper, with the mass fraction in halos ($p$),
    the halo bias ($b_h$), and the dust bias ($b_d$). We consider both real-space (r-space) 
    and redshift-space (z-space) cases.}
  \label{tab:1}
\end{table}

\section{Modeling the mass that is not observed}\label{sec:dust}
This section develops and tests a simple model for the dust, referred to as model (G), which can be generated based solely on the observed locations of the biased tracers. We compare it to model (W), which assumes that the dust follows the cosmic web. We use OT-G and OT-W to refer to analyses with these two models for the dust.  Model (W) is correct by definition, but impossible to implement in real data. On the other hand, (G) is relatively easy to implement.  While it is not accurate on small scales, we show below that OT-G accurately reconstructs the BAO feature in the biased-tracer field.  In Appendix~\ref{app:U}, we take a closer look at the dust model and its impact on OT, with the aim to understand its robustness.  In addition to treating model (G) in some detail, we also study a model (U), in which the dust is (wrongly) assumed to be uniformly distributed.  

To begin, Figure~\ref{fig:fields} displays the density fields of the dark matter, halos, and dust, obtained from the HADES cosmological simulations \cite{hades}.  The shading represents the mass density in units of the background:  $\rho/\bar\rho \equiv 1+\delta$ (recall that each halo is weighted by its mass).
Our first task is to use the (mass-weighted) halo distribution shown in the middle panel to guide the generation of a `dust' distribution which mimics that shown in the panel on the right, because our implementation of OT-reconstruction needs the sum of the two fields -- the total mass distribution -- to be as close to the panel on the left as possible.  

\subsection{A Wiener-filter linear bias model}
Motivated by the fact that the halo and dust samples appear to have similar spatial distributions, a simple model is to assume that fluctuations in the dust (which is not observed) are linearly proportional to those in the biased tracer (which is):  $\widehat{\delta_d} = w\,\delta_b$.  The constant of proportionality is obtained in the standard way, by minimizing $(\delta_d - \widehat{\delta_d})^2$ with respect to $w$, where $\delta_d$ is the true dust overdensity fluctuation.  This yields $\langle \delta_d\delta_b\rangle = w\,\langle\delta_b^2\rangle$ or the `Wiener filter' weight
\begin{equation}
    w = P_{db}(k)/P_{bb}(k),
    \label{eq:wiener}
\end{equation}
where $P_{ij}(k)$ are measured power spectra (i.e. $P_{ii}(k)$ includes a shot-noise contribution).  In addition, if 
 $r_{ij}\equiv P_{ij}/\sqrt{P_{ii} P_{jj}}$, 
then the Wiener filter estimator has $r_{d\widehat{d}} = r_{db}$ and 
 $P_{\widehat{d}\widehat{d}} = r_{db}^2\,P_{dd}$.  
Typically, $r_{db}\to 1$ as $k\to 0$, but it decreases as $k$ increases, suggesting that if we generate the dust using $w$ times the observed tracer field $\delta_b$, the result will be like smoothing the true dust field with $r_{db}$.  So it is reasonable to ask if this effective smoothing compromises the fidelity of the Optimal Transport reconstruction? In addition:  The Wiener filter requires knowledge of the $k$-dependence of $P_{db}(k)$ and $P_{bb}(k)$.  This is, in principle, a significant amount of cosmological model-dependent information, so how much does this compromise the model-independence of OT analyses?

\begin{figure}
    \centering
    \includegraphics[width=0.8\linewidth]{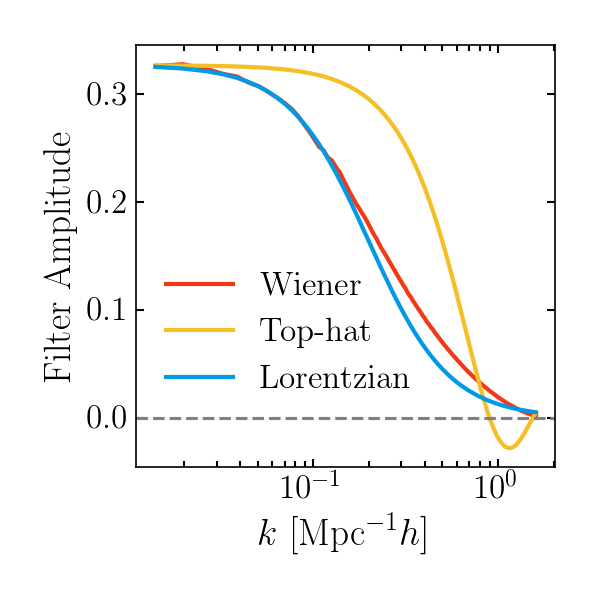}
    \caption{Simplest Wiener filter $P_{dh}/P_{hh}$ which, when applied to the halo field, provides the best linear model of the dust field (red). The other two curves show $(b_d/b_h)\, W(kR)$, with $W(x) = (1+x^2)^{-1}$ or $3j_1(x)/x$, and $R=5h^{-1}$Mpc.
    \label{fig:wiener}}
\end{figure}

To address this, we will use the mass-weighted halos as the biased tracer field,
so will set the subscript $b=h$ in what follows.  The solid red line in
Figure~\ref{fig:wiener} shows $P_{dh}(k)/P_{hh}(k)$.  Notice that it is
approximately constant at small $k$, after which it falls smoothly to zero at
larger $k$.  The two other curves show $(b_d/b_h)\,W(kR)$, 
where $b_d/b_h$ is the small $k$ limit of $\sqrt{P_{dd}(k)/P_{hh}(k)}$, 
$R=5h^{-1}$Mpc and
$W(x) = (1 + x^2)^{-1}$ or $3j_1(x)/x$ (labelled Lorentzian or TopHat). Thus, in this approximation, we model the dust as 
\begin{equation}
    \widehat{\delta_d}(k) = (b_d/b_h)\,W(kR)\,\delta_h(k).
    \label{eq:wienerd}
\end{equation}
We argue below that some of the $k$-dependence arises because the clustering of halos is mass-dependent, as is the halo-dust cross-correlation, and the formulation above ignores this.  The simpler analysis here highlights the fact that, to model the dust, we only need a single number, $b_d/b_h$, and an estimate of the scale $R$.  However, because 
\begin{equation}\label{eq:consistency}
    pb_h + (1-p)b_d = 1,
\end{equation}
we really need to know the mass fraction $p$ and bias factor $b_h$ of the observed tracers.  If the tracer masses are accurate, and the survey volume is known, then knowing $p$ is equivalent to knowing the background density $\Omega_m$.  And, since the clustering strength of the tracers is measured, knowing $b_h$ is equivalent to knowing the clustering strength of the full field, which is often parameterized by $\sigma_8$.  Thus, our linear bias model for generating the dust from the observed tracer field -- which we will refer to as model (G) -- must assume values for $\Omega_m$ and $\sigma_8$.

\begin{figure}
    \centering
    \includegraphics[width=0.48\textwidth]{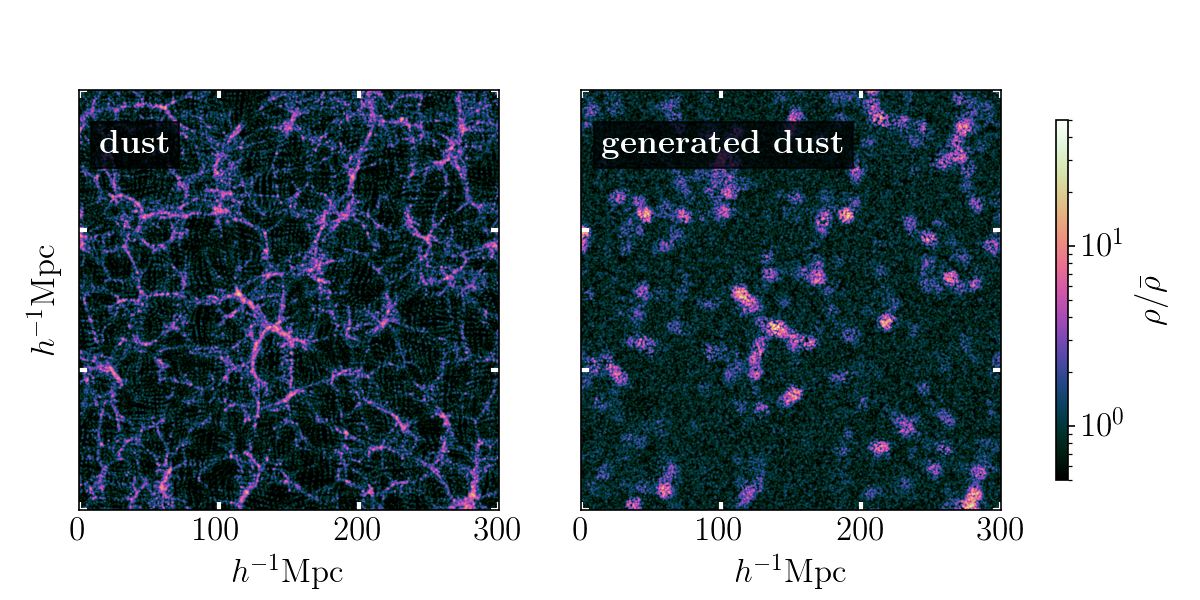}
    \caption{Comparison of original dust field (left, same as the right panel of Figure~\ref{fig:fields}) and that generated by our model (G) (right). 
    The generated field is like a smoothed version of the true dust field, as expected.
    \label{fig:gendust}}
\end{figure}

To implement this in practice, we Fourier transform the halo fluctuation field to obtain $\delta_h(k)$, `generate' $\delta_d(k)$ from it using equation~(\ref{eq:wienerd}), with $p$ and $b_h$ taken from Table~\ref{tab:1}, Fourier transform back, and then Poisson sample the resulting $1+\delta_d$ field using equal mass particles with mass $20m_p$.
Figure~\ref{fig:gendust} shows the result of this procedure when $R=5h^{-1}$Mpc.  
 (This is the smallest scale for which the `convex potential' assumption, on which OT is based, is expected to hold.)
The panel on the left shows the true dust field, and the one on the right shows our estimate of it, model (G), which was `generated' from the filtered halo field.  
We have checked that $r_{d\widehat{d}} = r_{dh}$ as expected, with small differences due to the fact that the $k$ dependence of equation~(\ref{eq:wienerd}) is not exactly correct.  
In addition, the generated field indeed appears to be a smoothed version of the true dust field (with $r_{dh}$ being an effective smoothing window).
Our next step is to see if this model for the dust is sufficiently accurate for our  reconstruction algorithm.  

\subsection{Dependence of OT-reconstruction on dust model}
\begin{figure}
    \centering
    \includegraphics[width=\linewidth]{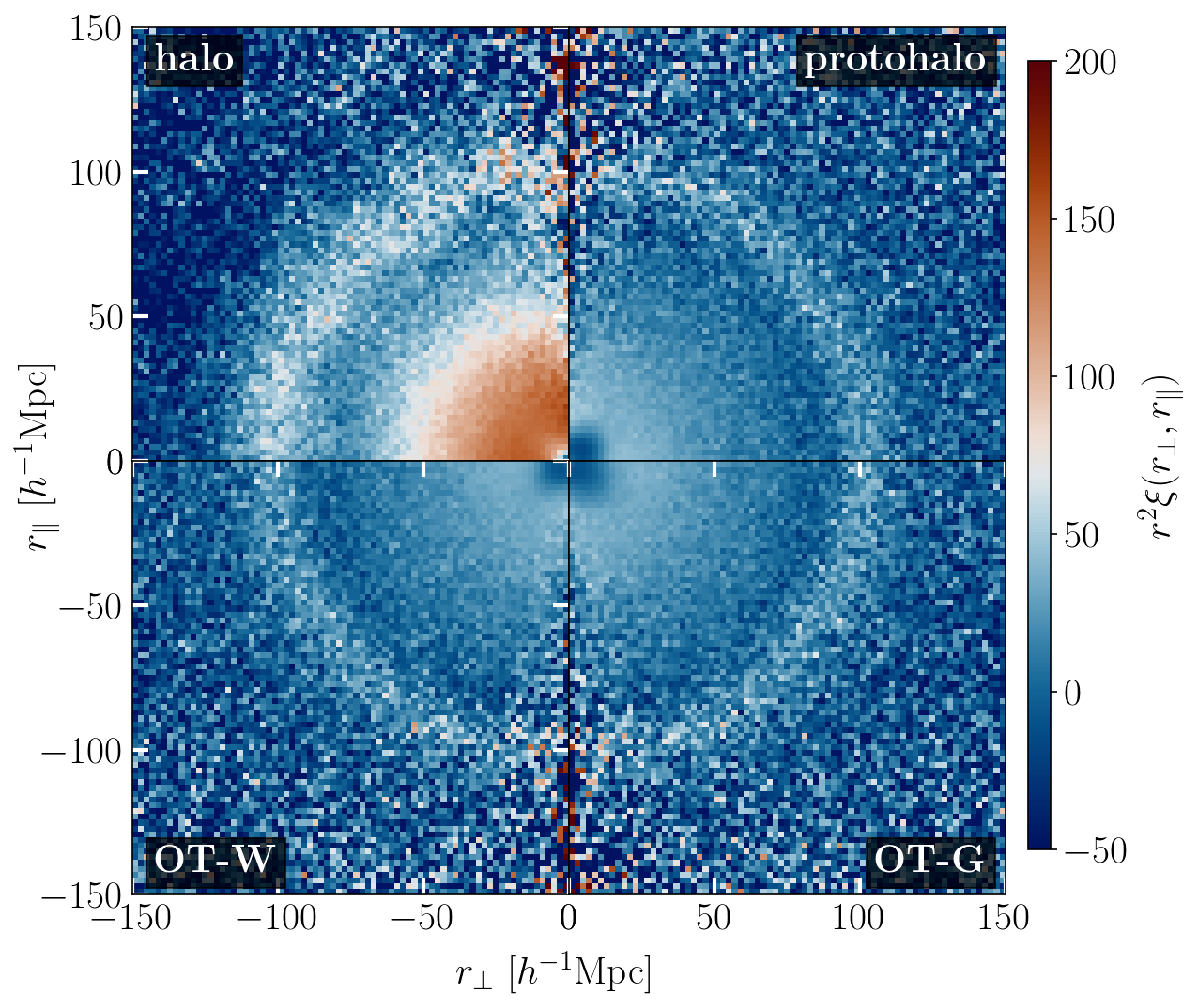}
    \caption{The real-space pair correlation function $\xi(r_{||},r_\perp)$ of
    the Eulerian halos prior to reconstruction (top left), Lagrangian protohalos (top right), and the OT-W and OT-G reconstructions (bottom).  In all panels, the BAO feature is apparent as a ring of radius $\sim 100h^{-1}$Mpc and width $\sim 10h^{-1}$Mpc, which is blurred-out in the top left panel and sharpened in the bottom panels.
    \label{fig:xi2D}}
\end{figure}

Figure~\ref{fig:xi2D} shows $r^2$ times the real-space pair correlation function $\xi(r_{||},r_\perp)$ of the halos, protohalos, reconstructed OT-G, and OT-W (clockwise from top left). In this format, the BAO feature is a ring of radius $\sim 100h^{-1}$Mpc. Prior to reconstruction, the width of the BAO feature in the evolved halo field is broad (top left); the ultimate goal of reconstruction is to sharpen this signal so it is like that in the initial Lagrangian protohalo distribution (top right). The BAO ring in the two post-reconstruction fields (OT-W and OT-G) is indeed closer to that of the protohalos.  As shown in Ref.~\cite{PRLhalos}, we expect OT-W to work well since it contains all the information of the underlying dark matter particles. However, the performance of OT-G is especially impressive because the dust model is simply a smoothed and scaled version of the biased tracer field.

\begin{figure}
    \centering
    \includegraphics[width=\linewidth]{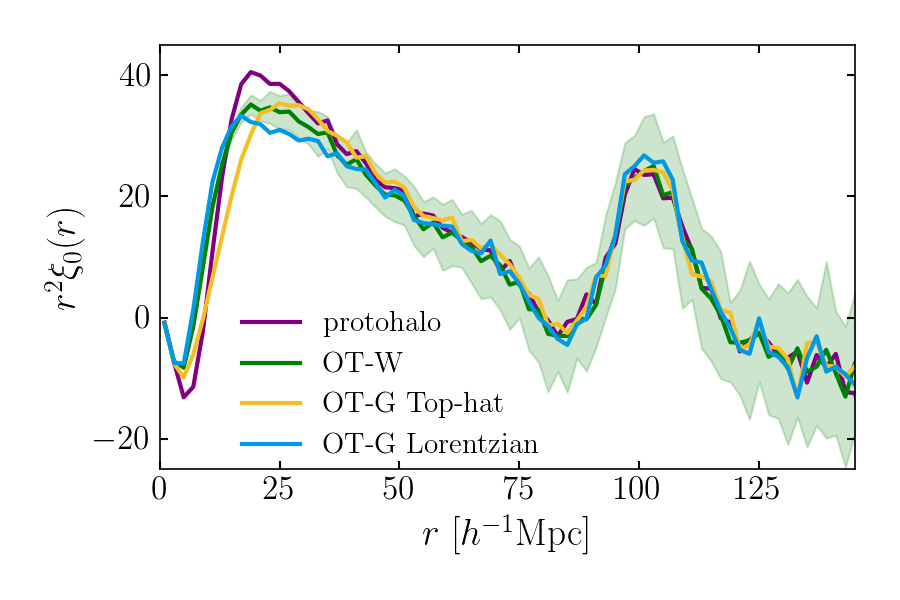}
    \caption{Monopole $\xi_0$ of the pair correlation function.  Purple curve shows the shape for the initial Lagrangian protohalos, while the other three show OT-reconstructed shapes:  green shows OT-W, and orange and blue show OT-G with dust generated using the Lorentzian and tophat filters shown in Figure~\ref{fig:wiener}.  The reconstructed BAO feature is robust to these differences:  post-reconstruction curves are in agreement with the protohalo curve down to $\sim 25 h^{-1}$Mpc.
    \label{fig:xi0}}
\end{figure}

The monopole of the protohalo, OT-W, and OT-G pair correlation functions is shown in Figure~\ref{fig:xi0}. Ref.~\cite{PRLhalos} argued that OT-W reconstructs the Lagrangian protohalo positions {\em extremely} accurately, so that the shape {\em and} amplitude of reconstructed correlation function provide novel constraints on the amplitude of the fluctuation field -- constraints that cannot be got from shape information alone. The remarkable agreement in both shape and amplitude between OT-G and protohalo's correlation function suggests that similar gains can be achieved using a simple dust model. 
However, recall that model (G) requires a guess about the halo mass fraction $p$ and the linear bias factor $b_h$.  Knowledge of $p$ implies an assumption about the value of matter density $\Omega_m$, and knowledge of $b_h$ implies an assumption about $\sigma_8$. Therefore, the dust model actually incorporates priors on $\Omega_m$ and $\sigma_8$. Although constraints on $\Omega_m$ are quite tight, $\sigma_8$ and hence the bias $b$ of the observed tracers, is only constrained to within 10 percent or so.  
In Appendix~\ref{app:modelG}, we study the dependence of OT-G on the guessed value for $b$, and show that variations in it do not affect the scale of the BAO feature, but do affect its amplitude. 

Before moving on, we note that we defined the dust field by smoothing the halos
with a spherical tophat of radius $R = 5h^{-1}$Mpc. We find that the reconstructed BAO feature in $\xi_0$ is insensitive to the
choice of $R$. However, the choice of $R$ does affect the clustering on smaller 
scales. This is not surprising.  In
essence, the actual protohalos have a fairly well defined exclusion radius,
below which $\xi\sim -1$.  This exclusion scale is not quite as sharp for OT-W
(because it doesn't reconstruct the protohalo positions and shapes exactly), and
gets increasingly blurred for OT-G as the smoothing scale $R$ is increased.
Since the smaller scale clustering is better reproduced if the smoothing scale
is smaller, $R=5h^{-1}$Mpc is our fiducial choice, even though a tophat of this
scale is not the shape singled out by our Wiener filter analysis
(Figure~\ref{fig:wiener}).

Although we have not done so here, we can build a more accurate model for the
dust which includes the fact that the tracers have different masses.  More
massive protohalos have larger exclusion radii, so including this may reduce the
discrepancies at small scales.  We could model $\widehat{\delta_d} =
\sum_h w_h\delta_h$ where the $h$ labels bins in halo mass.  Determining the
$w_h$ by minimizing $(\delta_d - \widehat{\delta_d})^2$ as before yields, via
the Sherman-Morrison formula,  
\begin{equation}
    w_h = \frac{(b_d/b_h)\,n_h b_h^2\,P_{mm}(k)}{1 + \sum_h n_h b_h^2 \,P_{mm}(k)}
\end{equation}
in the small-$k$ limit \cite[e.g. Ref.][]{cai2011}.  Implementing this requires
that we specify the mass fraction and bias factor in each bin.  If mass
estimates are in hand, and the clustering in each bin can be reliably measured
at least at small $k$, then this again only really requires that we specify
$\Omega_m$ and $\sigma_8$.  

Our approach here represents a simple first step towards modeling the dust, and demonstrates that even a simple model can yield good reconstructions. A number of more sophisticated models have been discussed in the literature \citep[e.g.][]{voronoi2020} with the recent work of \citep{AEdust} providing a particularly natural extension to our present approach. As it is likely that a detailed optimization of the dust model will also depend on the specific tracer population being considered (see the discussion around Fig.~\ref{fig:masscuts} below), we defer such investigations to future work.

\section{From redshift space distorted positions to initial conditions}\label{sec:zspace}
The results presented thus far have been in real-space. However, the challenge that remains is to account for the distortions that arise in redshift-space. 

\subsection{Anisotropic OT and effective domain}
In redshift space, the positions of galaxies appear to be displaced radially (along the line of sight) by an amount proportional to the radial component of the peculiar velocity. Therefore, the redshift position of an object located at the point $\bx$ in real space (Eulerian) can be expressed as
\begin{equation}
    \bx_{\rm rsd} = \bx + \frac{\hat{\bx} \cdot \bv}{a H} \hat{\bx},
\end{equation}
where $\bv$ represents the peculiar velocity in comoving coordinates, $\hat{\bx}$ is the unit normal vector in the direction of $\bx$, $a$ is scale factor, and $H$ is the Hubble parameter.  If $\bq$ represents the initial Lagrangian position, and $\bS \equiv \bx - \bq$ is the real-space displacement, then the redshift-space displacement is 
\begin{equation}\label{eq:rsd}
    \bS_{\rm rsd} = \bS + \frac{\hat{\bx} \cdot \bv}{a H} \hat{\bx}.
\end{equation}
This shows that to reconstruct real galaxy catalogs we must account for the velocity $\bv$ that causes redshift-space distortions. 

There are two distinct contributions to $\bv$: 
\begin{equation}
    \bv = \bv_{\rm coh}  + \bv_{\rm vir}.
\end{equation}
The first term represents coherent flows, which may be amenable to a perturbative treatment, and typically lead to a squashing along the line of sight \cite{kaiser1987, hamilton1992}.  The second term is random virial motions, which lead to `fingers-of-god' (FOGs) along the line of sight direction \cite{jackson1972}, but are confined to small nonlinear scales across it. The linear theory squashing depends on the cosmological model and weakly on the nature of the biased tracers, whereas the nonlinear FOGs are typically about 7 times longer than they are across. This means that FOGs can protrude into otherwise `linear' scales. So, as a first step, some redshift-space analyses attempt to identify these FOGs and `compress' them {\em prior} to reconstructing the field. In what follows, we will assume this has been done, so that $\bv = \bv_{\rm coh}$.  

\begin{figure*}
    \centering
    \includegraphics[width=\textwidth]{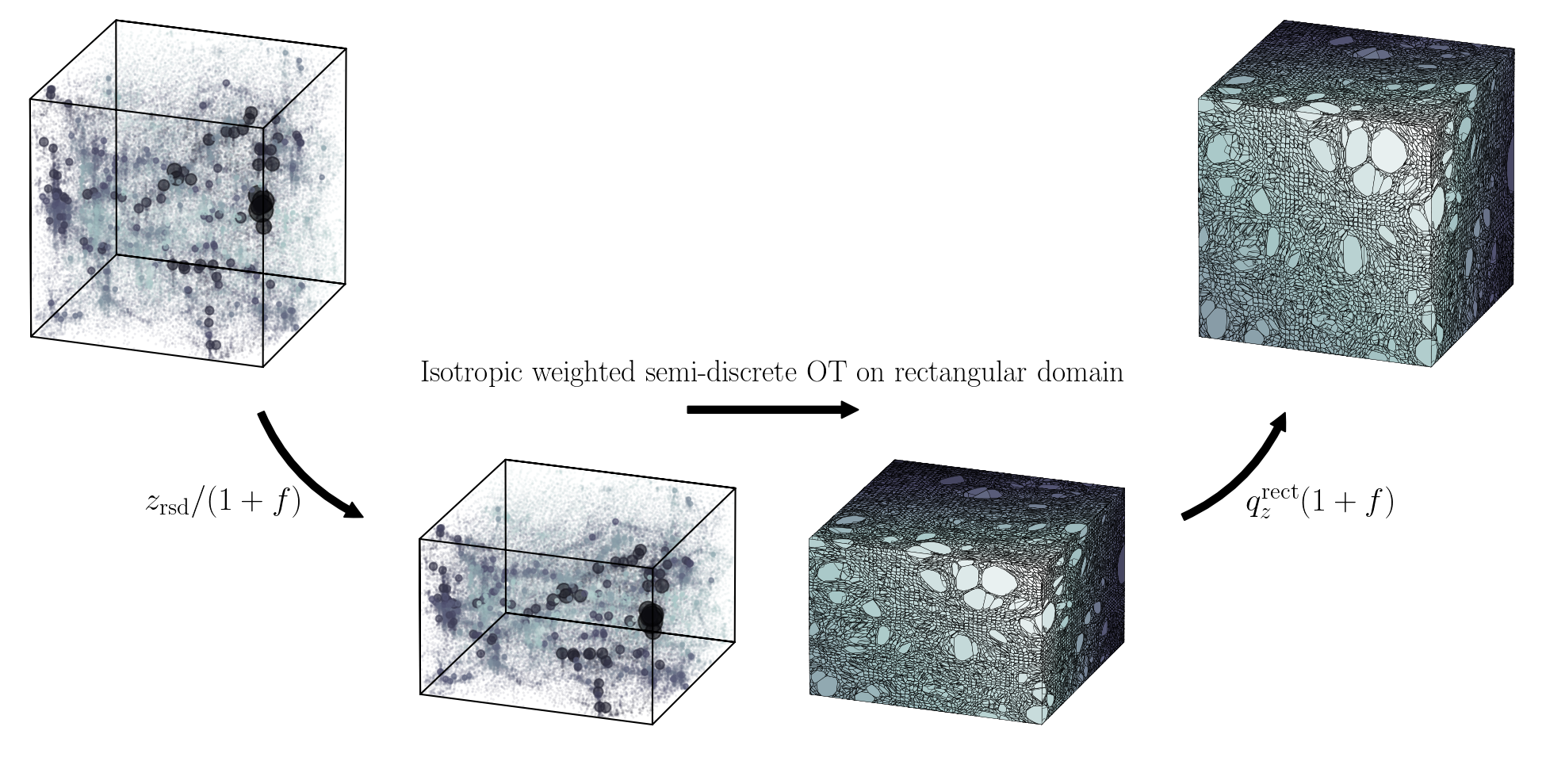}
    \caption{Schema showing how the semi-discrete algorithm reconstructs Laguerre cells from redshift space distorted positions, by   first transforming the domain, then reconstructing and finally transforming back.
    \label{fig:SDOT}}
\end{figure*}

In the Zeldovich approximation, the displacement vector from initial Lagrangian coordinates to evolved Eulerian coordinates is linearly proportional to the velocity vector:  
\begin{equation}
    \bv = afH\, \bS,
    \label{eq:vzel}
\end{equation}
where $f$ is the linear growth function. Substituting this into Eq.~\ref{eq:rsd} yields
\begin{equation}
    \bS_{\rm rsd} = \bS + f \,(\hat{\bx}\cdot \bS)\, \hat{\bx}.
    \label{eq:Szel}
\end{equation}
To connect redshift-space reconstruction to that in real-space, we use the fact that $\hat{\bx}\cdot\bS_{\rm rsd} = (1 + f)\, (\hat{\bx}\cdot \bS)$ to express the real-space displacement in terms of the redshift-space one:
\begin{equation}
    \bS = \bS_{\rm rsd} - \frac{f}{1+f} (\hat{\bx}\cdot\bS_{\rm rsd})\hat{\bx}.
\end{equation}
This makes 
\begin{equation}
    |\bS|^2 = |\bS_{\rm rsd}|^2 - \frac{f(2+f)}{(1+f)^2}\, (\hat{\bx}\cdot\bS_{\rm rsd})^2.
    \label{eq:cost}
\end{equation}
In the OT framework, $|\bS^2|$ is known as the `cost function' \cite{eur03}.  

We could write the cost schematically as the sum in quadrature of the displacement vector components perpendicular and parallel to the line of sight: $|S_{\rm rsd\perp}|^2 + |S_{\rm rsd ||}|^2/(1+f)^2$.  
This clearly treats the two components differently, and incorporating this would require modification of our semi-discrete OT algorithm to account for the anisotropy. 

However, in the plane-parallel, distant-observer approximation, with the $z$-coordinate as the line-of-sight direction, $\hat{\bx} = \hat{z}$, the quadratic cost function becomes
\begin{equation}
    |\bS|^2 = |\bS^{\rm rsd}_x|^2 + |\bS^{\rm rsd}_y|^2 + \frac{|\bS^{\rm rsd}_z|^2}{(1+f)^2}.
 \label{eq:costPP}
 \end{equation}
We can transform this into an isotropic problem by noting that 
\begin{widetext}

\begin{multline}
    \begin{pmatrix} 
    \bx^{\rm rsd}_x - \bq_x & \bx^{\rm rsd}_y - \bq_y & \bx^{\rm rsd}_z - \bq_z
    \end{pmatrix}
    \begin{pmatrix}
        1 & 0 & 0\\
        0 & 1 & 0\\
        0 & 0 & 1/(1+f)^2
    \end{pmatrix} 
    \begin{pmatrix} 
    \bx^{\rm rsd}_x - \bq_x\\ \bx^{\rm rsd}_y - \bq_y\\ \bx^{\rm rsd}_z - \bq_z 
    \end{pmatrix} \\ \equiv
    \begin{pmatrix} 
    \bx^{\rm rsd}_x - \bq_x & \bx^{\rm rsd}_y - \bq_y & \frac{\bx^{\rm rsd}_z - \bq_z}{1+f}
    \end{pmatrix}
    \begin{pmatrix}
        1 & 0 & 0\\
        0 & 1 & 0\\
        0 & 0 & 1
    \end{pmatrix}
    \begin{pmatrix} 
    \bx^{\rm rsd}_x - \bq_x\\ \bx^{\rm rsd}_y - \bq_y\\ \frac{\bx^{\rm rsd}_z - \bq_z}{1+f} 
    \end{pmatrix}.
\label{eq:metric2domain}
\end{multline}

\end{widetext}

The unequal entries on the diagonal of the matrix on the left indicate that the metric is not isotropic.  However, on the right hand side, the $z$-coordinates of both the redshift-space (Eulerian) and Lagrangian vectors have been rescaled by $1+f$, thereby changing the problem's domain, but, in this domain, the metric is isotropic.  This motivates the OT algorithm which is outlined in Figure~\ref{fig:SDOT}:  We rescale Eulerian positions by $1+f$ along the $z$-axis, run our isotropic OT reconstruction on this particle distribution in the anisotropic domain, and then rescale Lagrangian positions back.  Note that we do {\em not} require the final reconstructed and rescaled field to be isotropic.  
(In principle, our SD-OT algorithm weights each biased tracer by an estimate of its mass, since this mass is proportional to its Lagrangian volume.  Since we have changed the domain, we must rescale these weights by the same factor.  In practice, we always work with volume {\em fractions}, so this rescaling is not necessary.)

\subsection{Reconstructed displacements and cross-correlations}
\begin{figure}
    \centering
    \includegraphics[width=\linewidth]{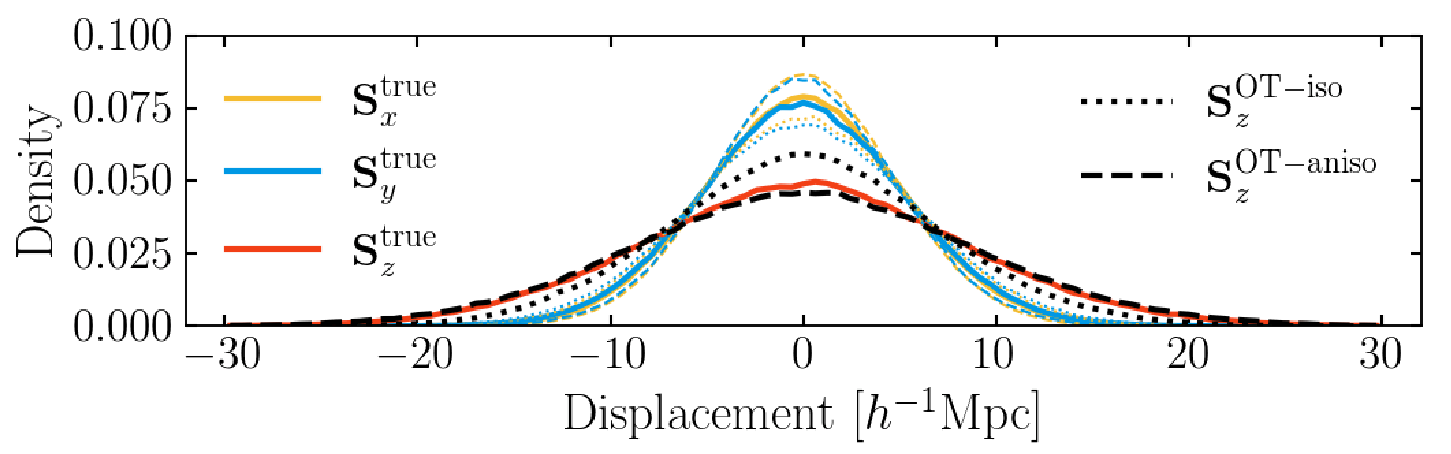}
    \caption{Distribution of displacements ($\bS_{\rm rsd}-\bq$) for our fiducial halo sample (solid) with those returned by the OT-Wz method outlined in Figure~\ref{fig:SDOT} (dashed), and those returned if one ignores the anisotropy in $\bS_{\rm rsd}$ (dotted).  
    \label{fig:displacement}}
\end{figure}

The displacements returned by this method depend on what we assume for the dust:  as before, we will consider two dust models, this time defined in redshift space (see Appendix~\ref{app:modelZ}), and we refer to the associated reconstructions as OT-Wz and OT-Gz.  To illustrate, Figure~\ref{fig:displacement} compares the true distribution of redshift-space distorted displacements with those returned by our `anisotropic' OT reconstruction and those it returns if we ignore the anisotropy altogether (i.e. we apply the real-space algorithm to the redshift space positions).  Our new method returns displacements in the $z$-direction that are much closer to the true values, although it produces slightly more peaked distributions for the components that are perpendicular to the line of sight.

\begin{figure}
    \centering
    \includegraphics[width=\linewidth]{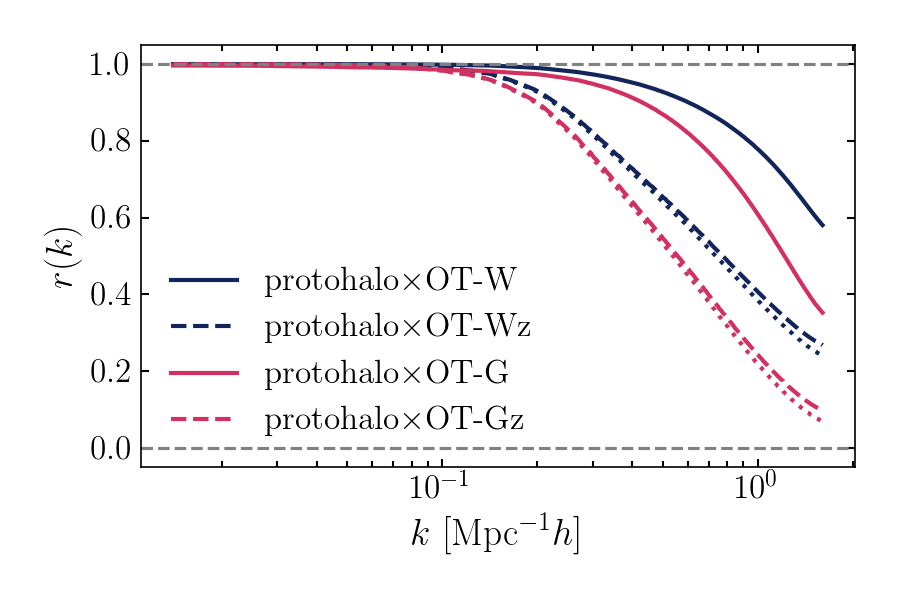}
    \caption{Cross-correlation coefficient $r(k)$ between the protohalo and reconstructed fields.  Solid and dashed curves show results which started from real and redshift space positions; the two curves of each type used different models for the dust (OT-Wz and OT-Gz).  Dotted curves show corresponding results if one treats the redshift space distorted positions as though they were true positions. 
    \label{fig:rkOT}}
\end{figure}

As another simple check of the fidelity of the OT-Wz and OT-Gz reconstructions, Figure~\ref{fig:rkOT} shows the cross-correlation coefficient between the protohalo and reconstructed fields.  To separate the impact of the dust model from that of the redshift-space anisotropies, the two solid curves show the real space results, OT-W and OT-G, and the dashed curves show the corresponding OT-Wz and OT-Gz.  Comparison of the two solid curves shows that although OT-G is worse than OT-W, it still has $r(k)\sim 1$ on BAO scales ($k\sim 0.1h$/Mpc). While the corresponding redshift space reconstructions are considerably worse at large $k$, they still have  $r(k)\sim 1$ at small $k$.  We argue below that OT-Gz remains good enough for BAO analyses.  

The dotted curves in Figure~\ref{fig:rkOT} show the result of ignoring the anisotropy all together, and simply treating the redshift space distorted positions as though they were the true positions.  They are remarkably similar to the dashed curves:  evidently, the OT reconstruction is not very sensitive to redshift-space distortions.  Since directions perpendicular to the line of sight are not distorted, it may be that the requirement of uniform/smooth initial conditions strongly constrains the extra displacements in the third direction that are required to undo the redshift-space distortions. 

\begin{figure}
    \centering
    \includegraphics[width=\linewidth]{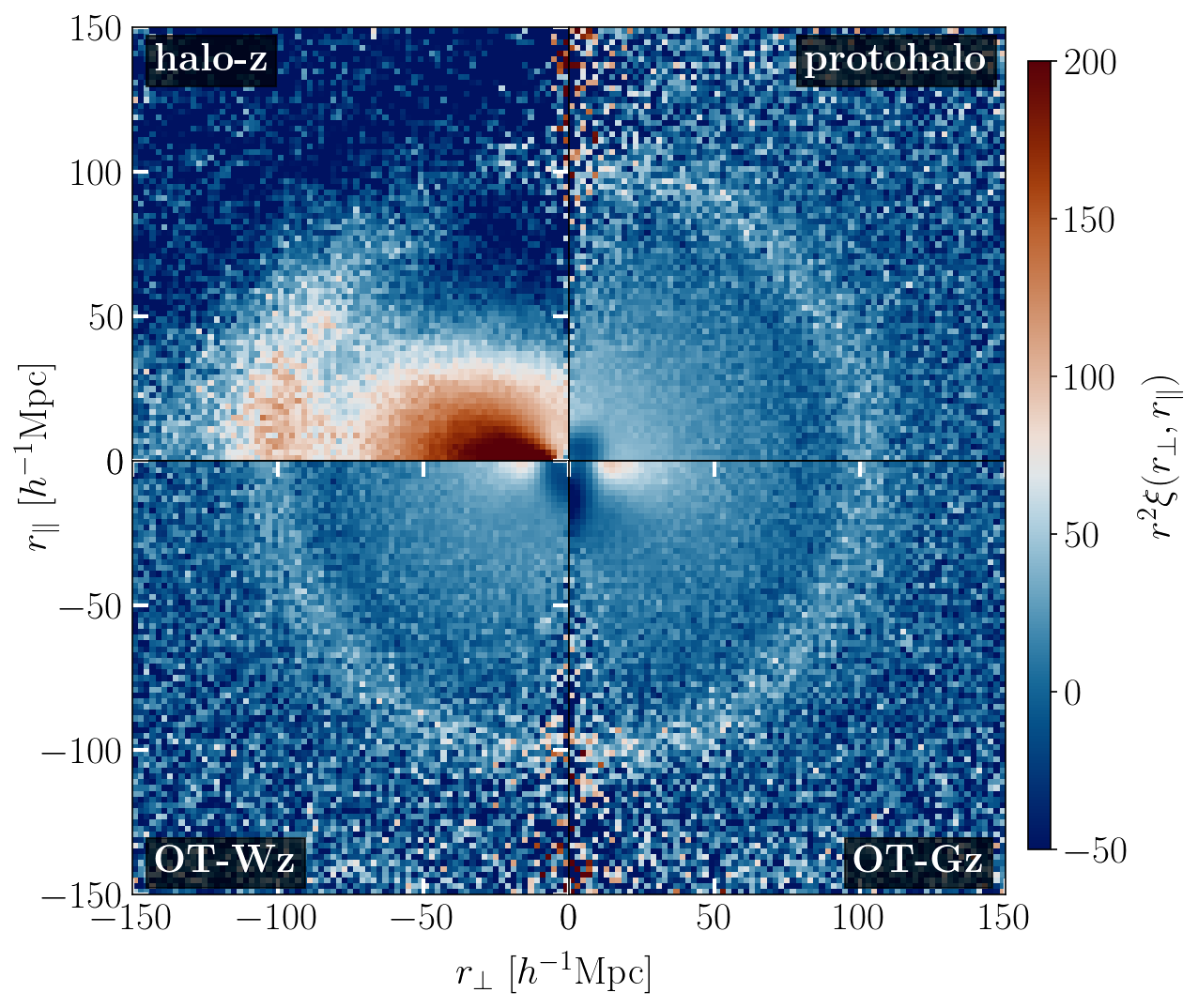}
    \caption{Same as Figure~\ref{fig:xi2D}, but now the top left panel uses Eulerian anisotropic redshift-space distorted positions.  Top right uses the real-space Lagrangian positions of the protohalos.  Bottom right and left show the result after running our new OT-reconstruction with models (Wz) or (Gz-tophat) for the dust in redshift-space.
    \label{fig:xi2D-zspace}}
\end{figure}

\subsection{Reconstructed pair correlations}
We now study the pair correlation function -- the analog of Figure~\ref{fig:xi2D}.  The left hand panel of Figure~\ref{fig:xi2D-zspace} shows the observed redshift-space distorted pair counts, as a function of separation along and across the line of sight:  the anisotropy -- a squashing along the line of sight -- is obvious.  The Lagrangian pair counts are in the next panel; they are isotropic (same as second panel of Figure~\ref{fig:xi2D}).  Our goal is to reconstruct this isotropic distribution starting from the positions which gave the distribution on the left.  The final two panels show the pair counts in the OT-Wz and OT-Gz fields (i.e., based on the approach of Figure~\ref{fig:SDOT}, but for two different models for the dust). 

\begin{figure}
    \centering
    \includegraphics[width=\linewidth]{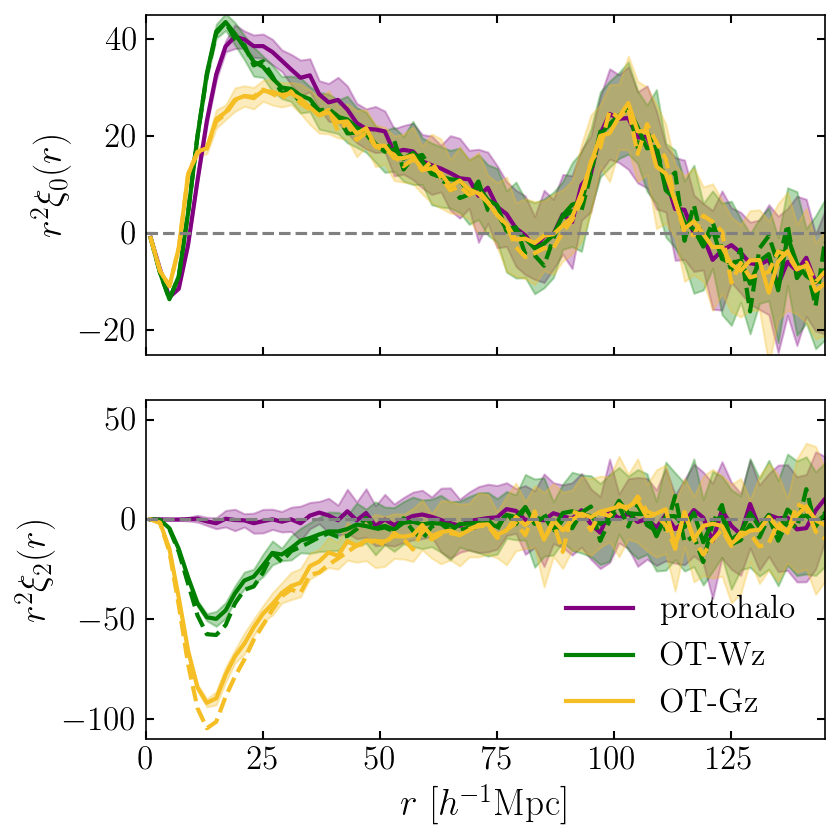}
    \caption{Comparison of the monopole (top panel) and quadrupole (bottom) of the pair correlation function of protohalo positions, and of the OT-Wz and OT-Gz reconstructed positions (purple, green and yellow).  The dashed curves in the bottom panel show the result of ignoring the fact that redshift-space distortions are anistropic, and simply treating the Eulerian positions as though they were in real-space.
    \label{fig:OTG-zspace}}
\end{figure}

Figure~\ref{fig:OTG-zspace} provides a more quantitative comparison.  The top and bottom panels show the monopole $\xi_0(r)$ and quadrupole $\xi_2(r)$ of the pair correlation function (sum the pair counts in Figure~\ref{fig:xi2D}, weighting each pair by ${\cal P}_\ell(\mu)$ where $\mu = r_{||}/r$ with $r = \sqrt{r_{||}^2 + r_\perp^2}$).  The top panel ($\xi_0$) should be compared with Figure~\ref{fig:xi0}.  The purple curves (for protohalos) in the two figures are the same, by definition; the green curves (dust model W) are very similar on BAO scales, where they are similar to the purple, but on smaller scales, OT-Wz produces stronger clustering than OT-W.  In contrast, OT-Gz produces weaker small scale clustering than OT-G, although it too works very well on BAO scales.  (The weaker small scale clustering is perhaps not surprising, given that the dust was defined using a smoothed field, so small-scale fluctuations have been removed.)

\begin{figure}
    \centering
    \includegraphics[width=\linewidth]{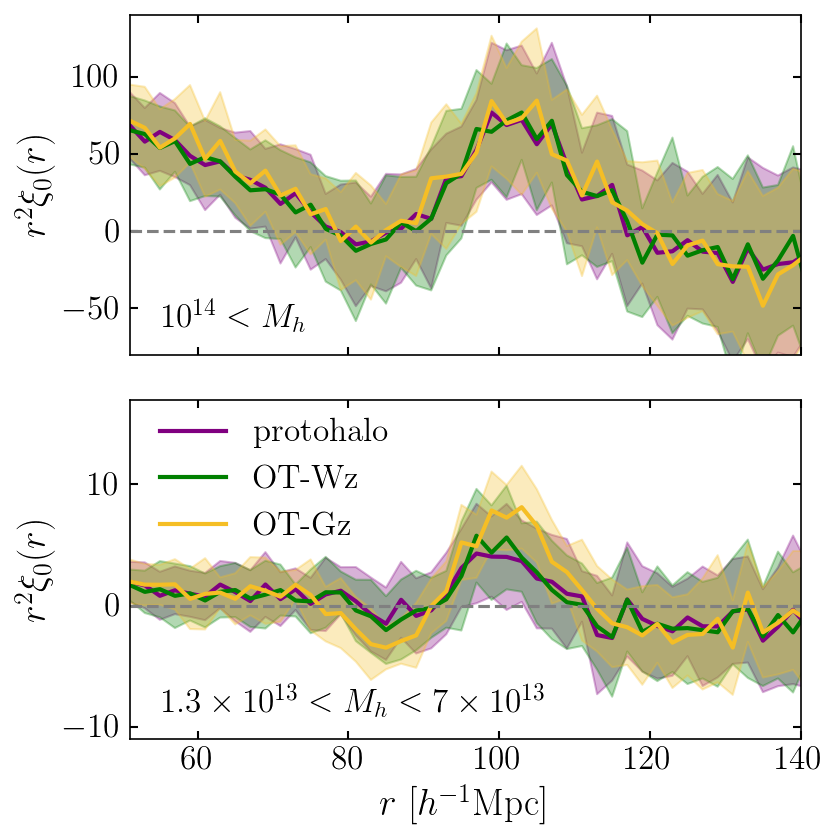}
    \caption{Same as top panel of Figure~\ref{fig:OTG-zspace}, but now showing the monopole for a high mass (top) and low mass (bottom) subset, chosen to have mass fraction $p\sim 0.1$ (see Table~\ref{tab:1}).
    \label{fig:masscuts}}
\end{figure}

Before we consider the quadrupole, Figure~\ref{fig:masscuts} shows the reconstructed $\xi_0$ in high- and low-mass subsamples.  These each have about half the mass density of the full halo sample, but very different bias factors (see Table~\ref{tab:1}).  For these smaller $p$ values modeling the dust is more important.  Nevertheless, both OT-Wz and OT-Gz reproduce $\xi_0$ of the actual protohalos extremely well, demonstrating that the accuracy of our methodology is not limited to a particular range in $b$ (or $p$).  

\begin{figure}
    \centering
    \includegraphics[width=\linewidth]{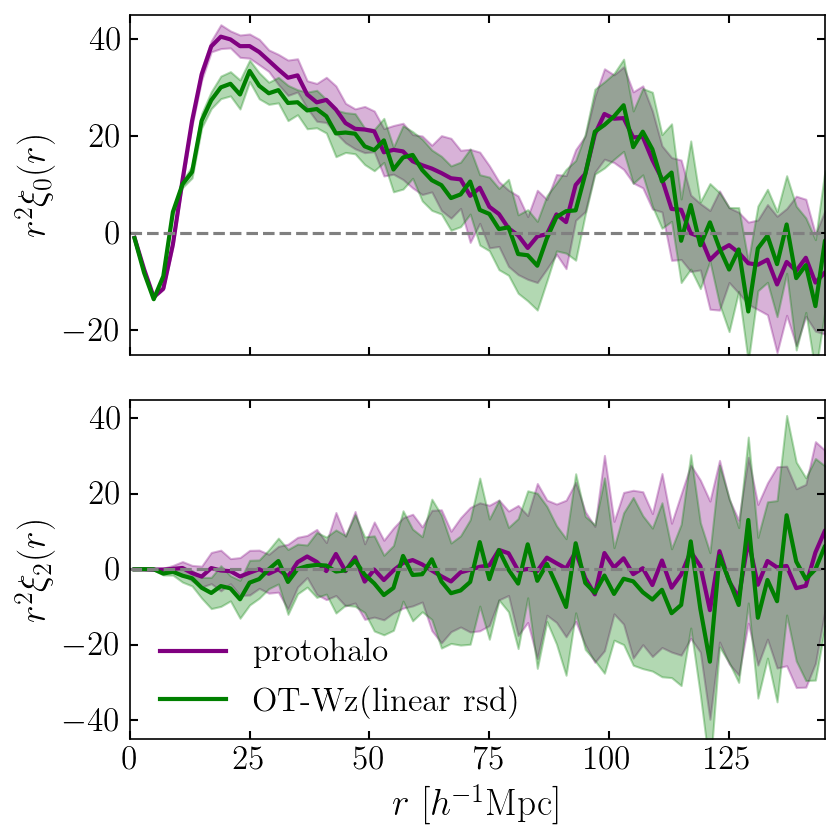}
    \caption{Reconstructed monopole and quadrupole in a model for which equations~(\ref{eq:vzel}) and~(\ref{eq:Szel}) are exact.  In this case, the reconstructed OT-Wz correctly removes the redshift-space anisotropies and returns a field with essentially zero quadrupole.   
    \label{fig:linrsd}}
\end{figure}

We turn now to the reconstructed quadrupole, shown in the bottom panel of Figure~\ref{fig:OTG-zspace}.  The true quadrupole (purple) is zero, of course, so it is very reassuring that both OT-Wz and OT-Gz also have $\xi_2\sim 0$ on BAO scales.  (The quadrupole of the left panel of Figure~\ref{fig:xi2D-zspace}, not shown, has $r^2\xi_2(r)\sim -75$, so the reduction is substantial.)  However, a significant (though reduced) quadrupole remains on smaller scales; using a better dust model helps (OT-Wz is closer to zero than OT-Gz).  To see if this is an artifact arising from our Zeldovich-motivated scaling, the dashed curves in this panel show $r^2\xi_2$ if we (incorrectly) treat the redshift-space positions as though they were in real space (i.e. we do {\em not} implement the domain rescaling step).  Comparison with the solid curves shows that the additional error this introduces is surprisingly small, but consistent with Figure~\ref{fig:rkOT} (where dotted and dashed curves were similar).  By this measure also, improving the dust model is more important than trying to account for the anisotropy.

Nevertheless, to understand the origin of the non-zero quadrupole, which is present even for OT-Wz, we have done the following test.  In the simulations, we know the actual real-space displacement vector $\bS$.  So, we used $f\bS$ instead of the actual $\bv$ to generate $\bS_{\rm rsd}$.  I.e., we ensure that equation~(\ref{eq:Szel}) is exactly correct.  We then run these redshift distorted positions through our OT-Wz pipeline.  Figure~\ref{fig:linrsd} shows the results.  The monopole is not substantially different, but, importantly, now $\xi_2 \approx 0$ on essentially all scales. We conclude that $\xi_2\ne 0$ results because equation~(\ref{eq:vzel}) is only approximately correct.  Using a less realistic dust model then amplifies the problem.

\subsection{Reconstructed higher order statistics}

\begin{figure}
    \centering
    \includegraphics[width=\linewidth]{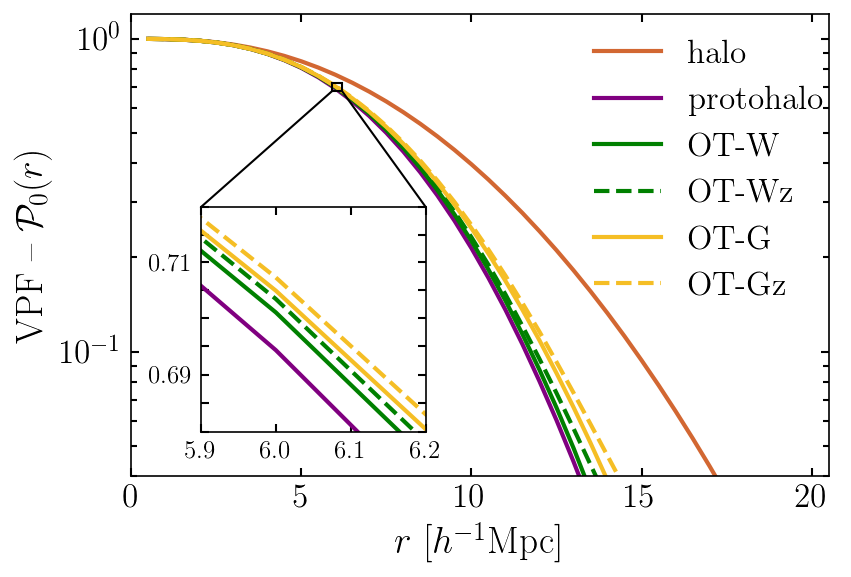}
    \caption{Void probability function for Eulerian halos in real space, Lagrangian protohalos, and the reconstructed OT-W, OT-G, OT-Wz, and OT-Gz halo positions.   
    \label{fig:vpf}}
\end{figure}

All our quantitative tests so far have used two-point statistics.  However, \cite{PRLhalos} showed that OT-W reconstructs the protohalo positions remarkably well, so it is likely that higher order statistics are also well-reproduced.  Although performing a systematic analysis of higher-order correlations is beyond the scope of this work, Figure~\ref{fig:vpf} shows the results of a simple test:  the void probability function -- the probability that a randomly placed sphere of radius $R$ is empty.  The VPF is known to depend on a specific combination of higher order moments, so provides a crude measure of whether or not the OT-reconstructions are accurate at higher order as well.  

The curve which is highest at large $R$ in Figure~\ref{fig:vpf} shows the VPF in of the evolved Eulerian halos, whereas the curve which is smallest is for the Lagrangian protohalos.  The number density of the two tracers is the same (by definition), so the difference between the two arises entirely because the protohalos are less strongly clustered.  (The VPF at $r\sim 15h^{-1}$Mpc is an order of magnitude smaller for the protohalos.)  All the other curves are very similar to the one for the protohalos, with small differences that are easily understood from the fact that OT-W is indeed very close to exact, OT-Wz reconstructs positions that are only slightly worse, OT-G positions are worse, and OT-Gz worse still.  Again, we see that modeling the dust well is more important than effects from redshift space distortions.

\section{Discussion} \label{sec:discussion}

We addressed two problems that are relevant for using Optimal Transport to reconstruct the cosmological distance scale that is encoded in the observed spatial distribution of biased tracers.  

The first has to do with modeling the `dust' that is not observed, but does affect the OT reconstruction.  Although Appendix~\ref{sec:dust} explores other models, for BAO purposes, we found that a Wiener filter-like model (equation~\ref{eq:wienerd}) which uses fluctuations in the observed field to generate fluctuations in the dust field works well (Figures~\ref{fig:gendust} and~\ref{fig:xi2D}).  We noted that knowing how much dust to generate is like assuming $\Omega_m$, and setting its clustering strength is like assuming $\sigma_8$.  The resulting reconstructions of the BAO feature are excellent (Figure~\ref{fig:xi2D}), down to scales of order $25h^{-1}$Mpc (Figure~\ref{fig:xi0}).  
Perhaps as importantly, we showed that it was possible to understand how mis-estimating either the dust fraction or clustering strength impact the OT reconstructions (equation~\ref{eq:bLagB} and Figures~\ref{fig:dust_bias} and~\ref{fig:bias-S}).  

The second has to do with accounting for the fact that the observed positions suffer from distortions induced by peculiar velocities along the line of sight.  This makes the reconstruction problem anisotropic.  However, if the peculiar velocities are aligned with the displacement (equation~\ref{eq:vzel}), then the anistropic OT reconstruction problem is simplified (equation~\ref{eq:cost}).  In the distant-observer limit, it can be transformed into an isotropic problem (equation~\ref{eq:metric2domain}) by rescaling the target domain, performing the Laguerre decomposition there, and then transforming back (Figure~\ref{fig:SDOT}).  Moreover, the dust model can also be extended to this case
(equation~\ref{eq:consistenzy}).  

Both the rescaling of the domain and the new dust model require additional cosmological information -- the growth factor $f$ -- which, in GR models, is approximately $\Omega_m^{4/7}$.  (Since we must assume $\Omega_m$ even for real space analyses, the new information is not the value of $\Omega_m$, but the fact that redshift space distortions depend on $f$ rather than some other quantity.)

Velocity and displacement are exactly aligned in the Zeldovich approximation -- i.e., to lowest order in Lagrangian perturbation theory -- so it should be reasonably accurate in real data.  We showed that this works well:  the cross-correlation coefficient between the OT-reconstructed and actual protohalo fields is close to unity down to $k\sim 0.2h$/Mpc (Figure~\ref{fig:rkOT}).  In addition, OT reconstructs the monopole $\xi_0$ and removes the quadrupole $\xi_2$ (i.e. restores isotropy) in the pair correlation function (Figure~\ref{fig:xi2D-zspace}) down to scales of order $25h^{-1}$Mpc (Figure~\ref{fig:OTG-zspace}).  

OT also reconstructs the void probability distribution rather well (Figure~\ref{fig:vpf}).  In principle, this enables cosmological constraints from statistics beyond two-point.  Our analyses of $\xi_2$ and of the VPF both suggest that there are more gains to be realized in a better modeling of the dust than in accounting for redshift space anisotropies (e.g. compare solid green with yellow vs dashed green in Figure~\ref{fig:OTG-zspace}).  Nevertheless, as it is robust to changes in the clustering strength and abundance of the tracer sample (Figure~\ref{fig:masscuts} and Table~\ref{tab:1}), even with its simple dust model, our method enables interesting OT analyses of BAO datasets.  

Although we have highlighted the fact that, in principle, our OT-Gz reconstructions must make assumptions about $\Omega_m, \sigma_8$ and $f$, in practice we have found that the reconstructions are not much degraded if we treat the redshift-space distorted positions as though they were the true real-space positions.  Neither the cross-correlation coefficient between the protohalo and reconstructed fields (dashed and dotted curves in Figure~\ref{fig:rkOT}) nor the reconstructed quadrupole (dashed and solid curves in Figure~\ref{fig:OTG-zspace}) are significantly degraded by ignoring redshift space distortions altogether.  This suggests that, at least on BAO scales, OT-Gz analyses do {\em not} really need to assume a value for $f$.  In addition, we showed that if one (wrongly) assumes that the dust is unclustered, then the scale of the BAO feature in the OT-Uz (rather than OT-Gz) reconstructed pair correlation function is unbiased (Figure~\ref{fig:xiU}).  Therefore, for studies which estimate the distance scale, the OT-Uz approach does not really need to assume $\sigma_8$ either.  This leaves $\Omega_m$ as the only cosmological parameter which must be assumed.  In future work, we will use our understanding of the dust model to assess the extent to which the dust model makes the OT-approach more cosmological model-dependent than originally hoped.

\bibliography{rsdOT.bib}

\appendix

\section{Models for the dust}\label{app:U}

We discuss why a model for the dust is necessary, and how this affects the reconstruction.  

\subsection{Unbalanced or Partial Optimal Transport}
Let us start from the fact that we know that if we have {\em all} the dark matter, then OT reconstructs the initial conditions in general, and the BAO feature in particular, very well \cite{PRLdm}.  Importantly, the `target' Lagrangian distribution, to which OT is directed to transport the mass, is simple to describe because it is unclustered.  If we only observe a biased subset of the mass, then the `target' distribution is complicated \cite{st1999}:  this, fundamentally, is why, in the cosmology context, a model for the dust is required.  

Such a model must specify $(a)$ how much dust, and $(b)$ how it is clustered.  
Consider $(a)$ first.  Suppose we only observe a biased subset of the mass, e.g. massive halos, but we have estimates of the mass of each tracer.  Suppose we also know that this biased subset of tracers accounts for a fraction $p$ of all the mass (that must be transported back to a uniform distribution).  There exist Optimal Transport formulations of such a problem, which go by the name of `Unbalanced' or `Partial' OT \cite{cpsv2015, sinkhorn2019, finiteTempOT}.  Typically, these modify the `cost function' or `constraints' part of the OT setup without explicit reference to question $(b)$.  In what follows, we discuss two crude models for $(b)$, which, once assumed, allow us to run OT as though it were `balanced'.  Of course, the result depends on what we choose for $(b)$, so it is not obvious that a single universal modification to the UOT cost that, presumably, accounts for $(a)$, will work for all choices $(b)$.  Stated differently:  it would be interesting to cast models (U) and (G) below in the UOT or POT frameworks.  E.g., it may be that model (U) is closest to the current POT literature, but showing this is beyond the scope of this work.

\subsection{Unclustered dust}
Before we offer a more quantitative discussion, the following qualitative argument, regarding the case in which the dust is assumed to be completely unclustered, may help highlight the issues.  

Suppose we only observe a biased subset of the mass:  e.g. halos down to some very low mass.  Then there would not be much missing mass, so what we assume for the `dust' cannot be that important.  In addition, low mass halos (i.e. those which include the missing mass) are not strongly clustered \citep{st1999}, so assuming they are completely unclustered is not too bad.  Next imagine that we increase the mass cut, so there is more missing mass.  Because the higher mass halos which now comprise some of the dust are more strongly clustered, we would expect the uniform dust model to fare worse.  To see how much worse, consider the opposite limit, in which the mass cut is so large that 99 percent of the mass is in dust.  In this limit, if we assume the dust is uniform, then the OT code is already starting from something that is close to uniform, so it only has to rearrange things so that the remaining 1 percent of the mass also ends up making the total uniform.  This means that the 1 percent does not move much (so $b_{\rm OT}$ of this 1 percent will be similar to $b_{\rm Eul}$); basically, OT will carve out an exclusion volume around these most massive halos, to account for the fact that they were $\sim 5$ times bigger at $z=\infty$ than they are today.  This is almost like undoing some of the smearing, but not exactly the same thing, since these massive halos also moved, but OT with uniform dust will get this wrong.  However, because they don't move much more than their Lagrangian size, getting it wrong is not too bad, so OT does manage to get a slightly sharper BAO feature.  (Of course, in this case, OT gets a better BAO feature, but does {\em not} reproduce the protohalo positions at all: $b_{OT} \sim b_{Eul}$ rather than $b_{\rm Eul}-1$.  On the other hand, if $b_{\rm Eul}\gg 1$ then  $b_{\rm Eul}-1 \sim b_{\rm Eul}$, so $b_{\rm OT} \sim b_{\rm Eul}$ is not so bad after all!).  So, its only in the regime between $\sim$30 percent and $\sim$90 percent that the uniform dust could end failing badly.  We will provide a more quantitative discussion of this uniform dust case shortly.  However, we hope that this discussion highlights the fact that one must know how much dust there is, and how it is clustered.

\subsection{Reconstructed positions or densities}

Although it was not the point of their work, the key to appreciating the role of dust is the work of \cite{mw1996}.  

Let $n_b$ denote the number density of objects (e.g. halos).  The expected number in a volume $V$ (e.g. a sphere) is $n_bV$.  The actual number in a randomly placed $V$ can fluctuate:  let $n_b V\,(1+\delta_b)$ denote this actual number, where $\delta_b$ fluctuates around a mean value of zero.  Let $n_b^{\rm Lag}$ denote the number density of protohalos.  Since there is one protohalo for each halo, $n_b^{\rm Lag} = n_b$.  The expected number of protohalos in an initial volume $V_{\rm Lag}$ is $n_b V_{\rm Lag}$; the actual number is $n_b V_{\rm Lag}\,(1+\delta_b^{\rm Lag})$, where $\delta_b^{\rm Lag}$ denotes the initial overdensity fluctuation (again, having mean 0).  \cite{mw1996} considered the case in which the initial $V_{\rm Lag}$ evolves into a final $V$, where $V$ is that volume which contains the mass that was initially within $V_{\rm Lag}$.  In this case, the protohalos in $V_{\rm Lag}$ become the halos in $V$, i.e., 
\begin{equation}
    n_b V\,(1+\delta_b) = n_b V_{\rm Lag}\,(1+\delta_b^{\rm Lag}).
\end{equation}
If the total mass (not the biased subset) was initially uniformly distributed, then it is natural to define the mass overdensity $\delta$ by $V_{\rm Lag}/V \equiv 1+\delta$:  the density in $V$ will be larger than in $V_{\rm Lag}$ if $V<V_{\rm Lag}$.  This makes the previous expression read 
\begin{equation}
    1+\delta_b = (1+\delta_b^{\rm Lag})(1+\delta);
\end{equation}
in effect, it provides a simple relation between the initial biased fluctuation $\delta_b^{\rm Lag}$ and the final one $\delta_b$.  Although $\delta_b^{\rm Lag}$ and $\delta_b$ refer to the biased tracers, $1+\delta$ refers to all the mass.  This shows that if one wishes to reconstruct the initial $1+\delta_b^{\rm Lag}$ from measurements of $1+\delta_b$ then one needs to know $1+\delta$.  
In the limit in which all fluctuations are small (typically large $V$ and $V_{\rm Lag}$), 
\begin{equation}
    \delta_b^{\rm Lag} = \frac{\delta_b - \delta}{1+\delta}\approx \delta_b - \delta.
\end{equation}
So, if $\delta_b^{\rm Lag} \approx b^{\rm Lag}\delta_{\rm Lag}$, 
    $\delta_b \approx b\delta$, 
    and $\delta\approx\delta_{\rm Lag}$, 
then
\begin{equation}
    b^{\rm Lag} = b - 1.
    \label{eq:bLbE}
\end{equation}
This `consistency relation' plays an important role in methods which seek to use both the scale and {\em amplitude} of the BAO feature to constrain cosmological model \cite{PRLhalos}.  However, if $1+\delta$ is not known, or is estimated incorrectly, then this simple relation may be violated.  

\begin{figure*}
    \centering
    \includegraphics[width=\textwidth]{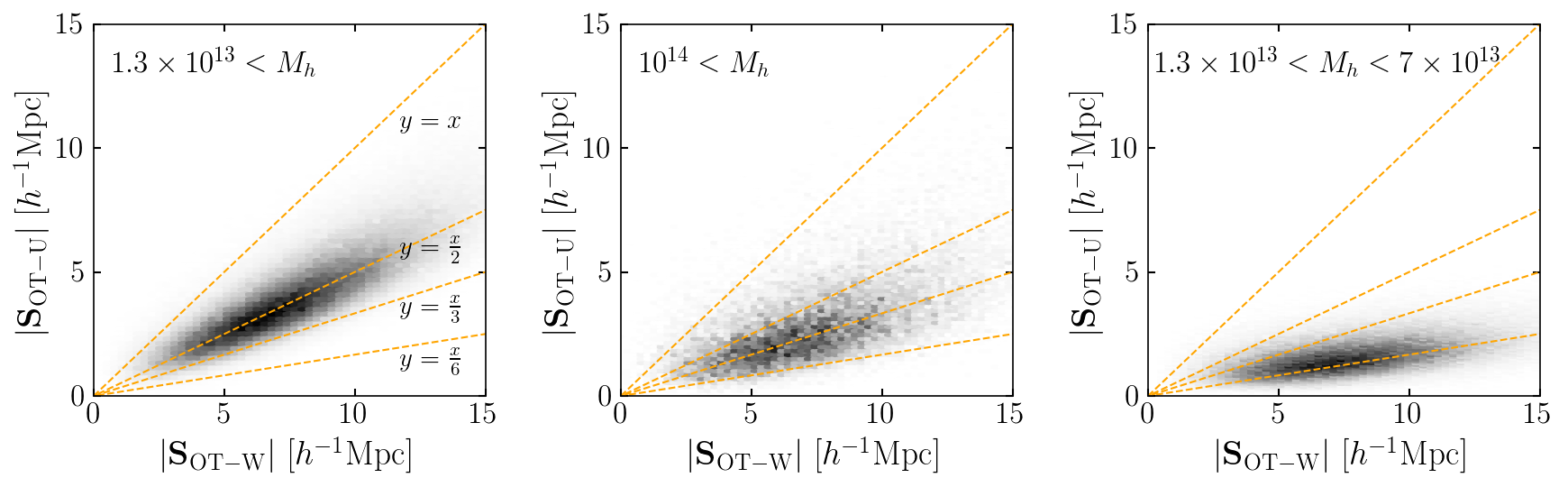}
    \caption{Comparison of OT-displacements $S_{\rm OT-U}$ in the uniform dust model with those from when the dust is drawn accurately to trace the cosmic web $S_{\rm OT-W}$, for a variety of choices of biased tracer.  Panels from left to right show results for all halos (mass fraction $p=0.22$), most massive or least massive halos selected to have the same mass fraction ($p=0.1$).  The mass cuts and associated bias factors are given in Table~\ref{tab:1}.  Equation~\ref{eq:SU} predicts that the slope of this relation should be $pb$.  
    \label{fig:uniform-S}}
\end{figure*}

Let $p$ denote the mass fraction in the observed tracers.  Then $1-p$ is the mass fraction in the rest, which we will refer to as `dust'.  If $\delta$ is the overdensity in the mass fluctuation field, then 
\begin{equation}    p\, (1 + \delta_b) + (1-p)\, (1 + \delta_d) = 1 +\delta
   \label{eq:consistency2}
\end{equation}
which means 
\begin{equation}
    p\, \delta_b + (1-p)\, \delta_d = \delta \quad {\rm so}\quad 
    \delta_d = \frac{\delta - p\,\delta_b}{1-p}.
\end{equation}
This shows that, unless $\delta_b = \delta/p$, the dust is clustered.  It also shows that, to estimate $\delta$ correctly, one must know both $p$ (essentially $\Omega_m$) and $\delta_d$ (which we will relate to clustering strength $\sigma_8$).  

\subsection{Connection to displacements}
Before we consider some explicit examples, it is interesting to study how the model for the dust impacts the displacements.  We noted before that 
\begin{equation}
    1+\delta = V_{\rm Lag}/V = (R_{\rm Lag}/R)^3
\end{equation}
If we write $R = R_{\rm Lag} - S$, where $S$ is the shift or displacement, then 
\begin{equation}
    1+\delta  = (1 - S/R_{\rm Lag})^{-3}.
\end{equation}
Hence, a mis-estimate of $\delta_d$ (such as those described in the previous subsections) will result in a mis-estimate of $1+\delta$, and hence a mis-estimate of the displacement $S$.  For $R_{\rm Lag}$ large compared to typical displacements, we expect
\begin{equation}
    1+\delta \approx 1 + 3 S/R_{\rm Lag} \quad {\rm so}\quad S\approx (\delta/3)\, R_{\rm Lag}.
\end{equation}
We will see shortly that this is a rather good way to think about how the OT-reconstruction depends on the model for the dust.

\subsection{Model U: Uniform dust}\label{app:modelU}
Suppose that one incorrectly assumes that the dust is uniformly distributed: i.e. one sets $\delta_d=0$.  Then one might define $\delta_{\rm U}$ from 
\begin{equation}
    p\, (1 + \delta_b) + (1-p) = 1 +\delta_{\rm U}
    \quad {\rm so}\quad 
    p\,\delta_b = \delta_{\rm U}
\end{equation}
and one would reconstruct 
\begin{equation}
    1+\delta_b^{\rm Lag-U} = \frac{1+\delta_b}{1 +\delta_{\rm U}}
     = \frac{1+\delta_b}{1 + p\delta_b}
\end{equation}
making 
\begin{equation}
    \delta_b^{\rm Lag-U} \approx \delta_b\, (1-p).
    \label{eq:dLagU}
\end{equation}
Note that this violates equation~(\ref{eq:bLbE}) if $pb\ne 1$.  
In general, the dependence on $b$ and $p$ quantifies the discussion at the start of this Appendix.
E.g., if $p\to 1$ then $\delta_b^{\rm Lag-U}\to 1$, since the OT algorithm tries to make the protohalos define a uniform field.  If $p\to 0$ then $\delta_b^{\rm Lag-U} \to \delta_b$ since then the mass is dominated by the dust which is already uniform, so the OT algorithm doesn't need to move the biased tracers much, meaning their spatial distribution is hardly modified.  

\begin{figure}
    \centering
    \includegraphics[width=\linewidth]{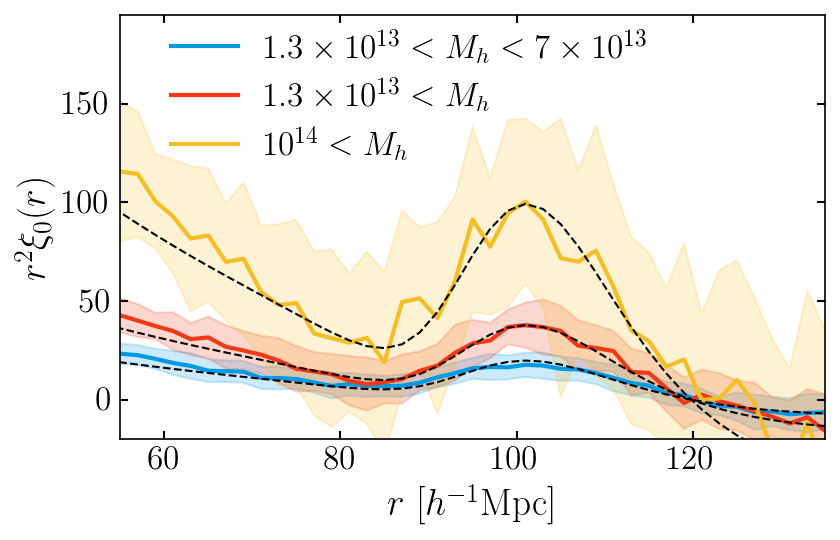}
    \caption{Dependence of OT-U reconstructed correlation function on the abundance $p$ and clustering strength $b$ of the biased tracers.  Smooth dashed curves show $b_{\rm U}^2\,\xi_{\rm Lin}(r)$ with $b_{\rm U}=b\,(1-p)$ (c.f. equation~\ref{eq:dLagU}) and $b$ and $p$ appropriate for the stated mass cuts (see Table~\ref{tab:1}).
    \label{fig:xiU}}
\end{figure}

Figure~\ref{fig:xiU} shows the reconstructed pair correlation function for the three halo samples described in Table~\ref{tab:1}: our fiducial halo sample, for which $p=0.22$ and $b=2$; 
 a more massive sample for which $p\approx 0.1$ but $b=3.19$ is larger; 
 and a less massive sample for which both $p$ is again smaller (0.11) but now $b$ is smaller (1.42).
 The dashed curves show $b_{\rm U}^2\,\xi_{\rm Lin}(r)$ with $b_{\rm U}=b\,(1-p)$ as given by equation~(\ref{eq:dLagU}); they describe the measurements extremely well.

We now turn to the displacements.  In this uniform dust model they should satisfy 
\begin{equation}
 S_{\rm U} = (\delta_{\rm U}/3)\,R_{\rm Lag} = (\delta_{\rm U}/\delta)\,S 
           = (p\delta_b/\delta)\,S \approx pb\,S.
 \label{eq:SU}
\end{equation}
i.e., if $pb < 1$ then $S_{\rm U}$ will be smaller than the true displacement, so the reconstructed $\delta_b^{\rm Lag-U}$ will be closer to $\delta_b$ than it should be.  Figure~\ref{fig:uniform-S} shows that this works rather well for all three cases, which have $pb=0.44, 0.3$ and $0.15$ for the fiducial, massive and low-mass halo samples.   

We noted in the main text that knowledge of $p$ and $b$ imply prior knowledge of $\Omega_m$ and $\sigma_8$.  This uniform dust model only uses $p$, so does not require knowledge of $\sigma_8$.

\subsection{Model G: Dust as linearly-biased version of tracer}\label{app:modelG}
Alternatively, suppose one wishes to model $\delta_d$ as a multiplicative factor times $\delta_b$:  
$\delta_d = \delta_b\,(\delta/\delta_b - p)/(1-p)$.   
To do this correctly, one must know $p$ and $\delta_b/\delta$.  If $\delta$ is not known, then the ratio $\delta_b/\delta$ must be guessed.  Suppose we guess that this ratio is a constant $B$.  Then we will set 
\begin{equation}
    \delta_d = \frac{\delta_b}{B}\frac{1 - pB}{1-p} 
\end{equation}
and we will mis-estimate $1+\delta$ as 
\begin{equation}
    1+\delta_{\rm B} \equiv p(1+\delta_b) + (1-p)(1+\delta_d) 
    = 1 + \frac{\delta_b}{B}.
\end{equation}

\begin{figure}
    \centering
    \includegraphics[width=\linewidth]{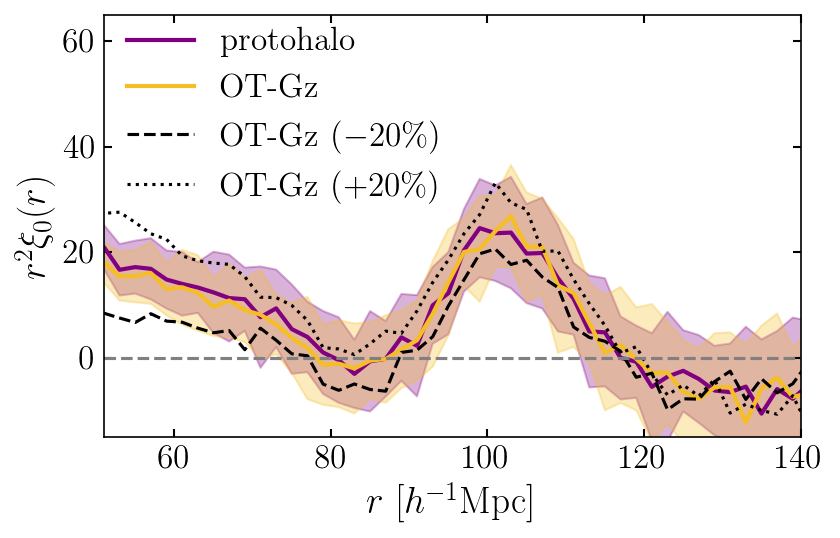}
    \caption{Dependence of OT-G on the assumed clustering strength of the biased tracers:  over/under estimating the Eulerian bias results in a reconstructed field in which the clustering is also too strong/weak.  
    \label{fig:dust_bias}}
\end{figure}

The reconstructed biased tracers will be 
\begin{equation}
 1+\delta_b^{\rm Lag-B} = \frac{1+\delta_b}{1 +\delta_{\rm B}} = \frac{1+\delta_b}{1 +\delta_b/B}
\end{equation}
so
\begin{equation}
 \delta_b^{\rm Lag-B} 
     = \frac{\delta_b - \delta_b/B}{1 + \delta_b/B}
     \approx \delta_b\,(B - 1)/B
\end{equation}
If $B=b$ then this correctly becomes $\delta_b-\delta$, but not otherwise.   
In the linear bias approximation this would become 
\begin{equation}
 b^{\rm Lag-B}
                = (b/B)\,(B-1)= (b - 1) + (1-b/B);
 \label{eq:bLagB}
\end{equation}
if $B>b$ then $b^{\rm Lag-B} > b^{\rm Lag-b}=b-1$; otherwise it will be too small.  
Figure~\ref{fig:dust_bias} shows that this is in qualitative agreement with the trends we see in the OT-reconstructed fields.  

\begin{figure}
    \centering
    \includegraphics[width=\linewidth]{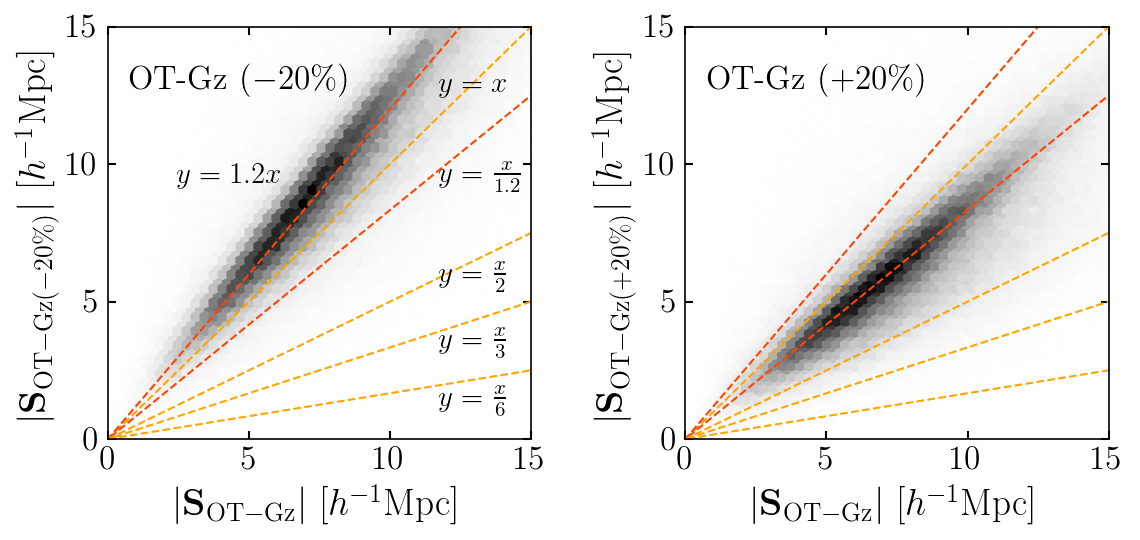}
    \caption{Same as left-hand panel of Figure~\ref{fig:uniform-S}, except that now the positions of tracers were given in redshift space, and the dust was assumed to be linearly proportional to the tracers, which were the full halo sample (equation~\ref{eq:delta_dz}).  Left and right hand panels show the effect on the displacements if the bias of the tracer sample is under- or over-estimated by 20\%.  Equation~(\ref{eq:SB}) predicts the slope of this relation to be given by the ratio of the true to estimated redshift-space distorted bias factors:  if $b$ and $B$ and the true and estimated bias factors, then their redshift space distorted values are $(b+f/3)/(B+f/3)$.
    \label{fig:bias-S}}
\end{figure}

The associated displacements should satisfy 
\begin{equation}
 S_{\rm B} = (\delta_{\rm B}/\delta)\,S = (\delta_b/B\delta)\,S\approx (b/B)\,S.
 \label{eq:SB}
\end{equation}
This shows that if $B>b$ (one has overestimated the strength of the bias of the observed tracers) then one will underestimate the displacement.  As a result, one will not move the biased tracers all the way back to their initial positions, which is why $\delta_b^{\rm Lag-B}$ will end up being closer to $\delta_b$ than it should be.  Figure~\ref{fig:bias-S} shows that this is in good qualitative agreement with the actual dependence of the OT-displacements on $B$.  

If both $p$ and $b$ are guessed wrong, so $\delta$ is also guessed wrong, and we used upper case to denote the guessed values, then requiring $p\,(1+b\delta) = P\,(1 + B\Delta)$ makes 
$\delta_b^{\rm Lag-B} 
     = (\delta_b - \Delta)/(1 + \Delta)
     \approx (P/p - 1)\, (1 + \Delta) + (P/p)(B-1)\Delta$.
In this case, it is useful to distinguish between $p$ being wrong because the mass estimates of the biased tracers are wrong versus assuming the wrong $\Omega_m$.  We do not pursue this further here, but note that this can, in principle, be done.

Before we end this section, it is worth making the following point.  
We noted that $b^{\rm Lag-B}$ of equation~(\ref{eq:bLagB}) would be `too large' if $B>b$.  
While this is true, if $b$ is not known, then how will we know if $b^{\rm Lag-B}$ is too large or too small?  
One might have hoped that we could use the fact that the square-root of the ratio of the Eulerian and reconstructed correlation functions should equal $b/(b-1)$ (cf. the consistency relation of equation~\ref{eq:bLbE}).  So, what if we make this ratio using our guessed value of $B$?  It will be 
\begin{equation}
    \frac{\delta_b}{\delta_b^{\rm Lag-B}} 
    = \delta_b\,\frac{1 + \delta_b/B}{\delta_b - \delta_b/B}
    \approx \frac{B}{B-1}
\end{equation}
if $\delta_b/B\ll 1$ (as is likely on large scales for realistic values of $b$ and $B$).  Since this ratio has the right structure, there is no way to know, at least from large/linear scale clustering, that we have guessed the wrong $B$.

\subsection{Dust in redshift space}\label{app:modelZ}
In redshift space, on large scales where linear theory applies, 
\begin{equation}
 \delta_b\to (b + f\mu^2)\,\delta
\end{equation}
\citep{kaiser1987}; this just expresses the fact that the displacement vector which relates an object's initial and final positions is proportional to (i.e. points in the same direction) as its present day velocity, with a constant of proportionality that is the same for all objects.  In this approximation,  equation~(\ref{eq:consistency}) becomes 
\begin{equation}
   p(b + f\mu^2) + (1-p) (b_d + f\mu^2) = pb + (1-p)\,b_d + f\mu^2 .
   \label{eq:consistenzy}
\end{equation}
For the correct $b_d$ we know that $pb + (1-p)\,b_d = 1$, 
so the monopole average over all $\mu$ is 
\begin{equation}
   p (b + f/3) + (1-p) (b_d + f/3) = 1 + f/3 .
\end{equation}
This, then, is the appropriate generalization of equation~(\ref{eq:consistency}), after which the same discussion regarding the impact of mis-estimates of $p$ or $b$ on the reconstructed field follows.  

In practice, the redshift-space quantity which is more familiar is the monopole average of 
 $(b + f\mu^2)^2\to b^2 + 2bf/3 + f^2/5$ 
\cite[e.g.][]{kaiser1987}.  If we call this $b_{\rm K}^2$ then, for large positive $b$, $b_{\rm K}\approx b + f/3$, so it can be used instead to determine 
\begin{equation}
 b_{dz} = (b_d + f/3)\approx (1 + f/3 - p\,b_{\rm K})/(1-p).
\end{equation}
and hence 
\begin{equation}
    \delta_{dz} = \frac{\delta_{bz}}{B_{\rm K}}\frac{1 + f/3 - pB_{\rm K}}{1-p} .
 \label{eq:delta_dz}
\end{equation}
But in general, the appropriate quantity is $b+f/3$.  

We end this section with the following observation.  Evidently, if only redshift-space distorted quantities are available, then, to model the dust, one needs to also know $f$.  In LCDM, $f\approx \Omega_m^{4/7}$, and since we have already made the point that knowing $p$ means knowing $\Omega_m$, there is a sense in which guessing $f$ does not really add new information (in the LCDM family of models).

\subsection{Robustness to mis-estimates of mass}\label{sec:LNmass}
\begin{figure}
 \centering
 \includegraphics[width=1\hsize]{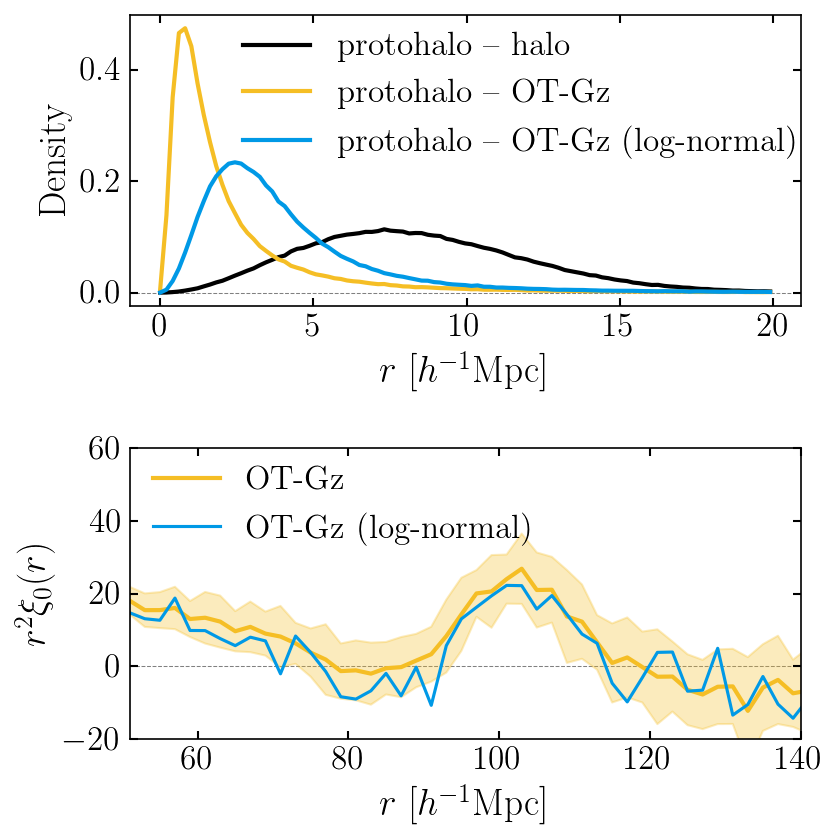}
 \caption{\label{fig:errm} Top: Distribution of displacements between Eulerian and Lagrangian positions (black), and OT and Lagrangian positions (colored).  If the masses of the biased tracers are known perfectly (green), OT reconstructs the positions typically to within about a Mpc; if halo mass estimates are noisy, and follow a log-normal distribution centered on the true mass, with unit rms (purple), then the quality of the reconstruction seems rather robust and is degraded only very slightly. Bottom:  Effect of log-normal errors, with $\sigma_M=1$, on $\xi_{\rm OT}$.
  }
\end{figure}

We have used a Log-normal model to quantify how mis-estimating the mass propagates into our analysis.  For the $i$th object we set $M_{{\rm est},i} = M_{{\rm true},i}\,\exp(\sigma_M g_i - \sigma_M^2/2)$, with $g_i$ drawn from a Gaussian distribution having zero mean and unit variance.  This makes the ratio $M_{{\rm est},i}/M_{{\rm true},i}=1$ on average, with rms around this mean of $\sqrt{\exp(\sigma_M^2)-1}$: larger values of $\sigma_M$ yield larger mis-estimates.  Figure~\ref{fig:errm} shows that even $\sigma_M=1$ only degrades the reconstructed positions by about 1-2 Mpc; this is not enough to compromise the improvements in the BAO distance scale such as those discussed in Ref.\cite{PRLhalos}.

\end{document}